\documentclass[aps,prl,reprint,superscriptaddress]{revtex4-1}

\usepackage{amsmath}
\usepackage{amsfonts}
\usepackage{textcomp}
\usepackage{graphicx}
\usepackage{epstopdf}

\newcommand{\eref}[1]{(\ref{#1})}

\begin{document}


\title{Spontaneous Chiral Symmetry Breaking in Planar Polarized Epithelia}


\author{Jeremy Hadidjojo}
\affiliation{Department of Physics, University of Michigan, Ann Arbor, Michigan, 48109-1040, USA}
\author{David K. Lubensky}
\affiliation{Department of Physics, University of Michigan, Ann Arbor, Michigan, 48109-1040, USA}
\email[]{dkluben@umich.edu}


\date{\today}

\begin{abstract}

Most animal body plans have some degree of left-right asymmetry.  This chirality at the tissue and organ level is often assumed to originate from the intrinsic handedness of biological molecules. How this handedness might be transferred from molecules to tissues during development, however, is not well understood. Here we explore an alternative paradigm where tissue chirality results from spontaneous symmetry breaking at the cellular scale, with molecular chirality acting only as a weak bias that ensures that one handedness predominates over the other. Specifically, we show that systems capable of generating planar polarity, found in many epithelial tissues, can also generically break left-right symmetry, and we identify the key interaction parameters that must be varied to access the chiral phase.  In addition to a chiral polar phase corresponding to one found in liquid crystal films, a two-dimensional chiral nematic phase with no liquid crystal analog is also possible.  Our results have clear implications for the interpretation of many mutant phenotypes, especially in certain \textit{Drosophila} epithelia.

\end{abstract}


\maketitle

\begin{figure}
	\includegraphics[scale=0.6]{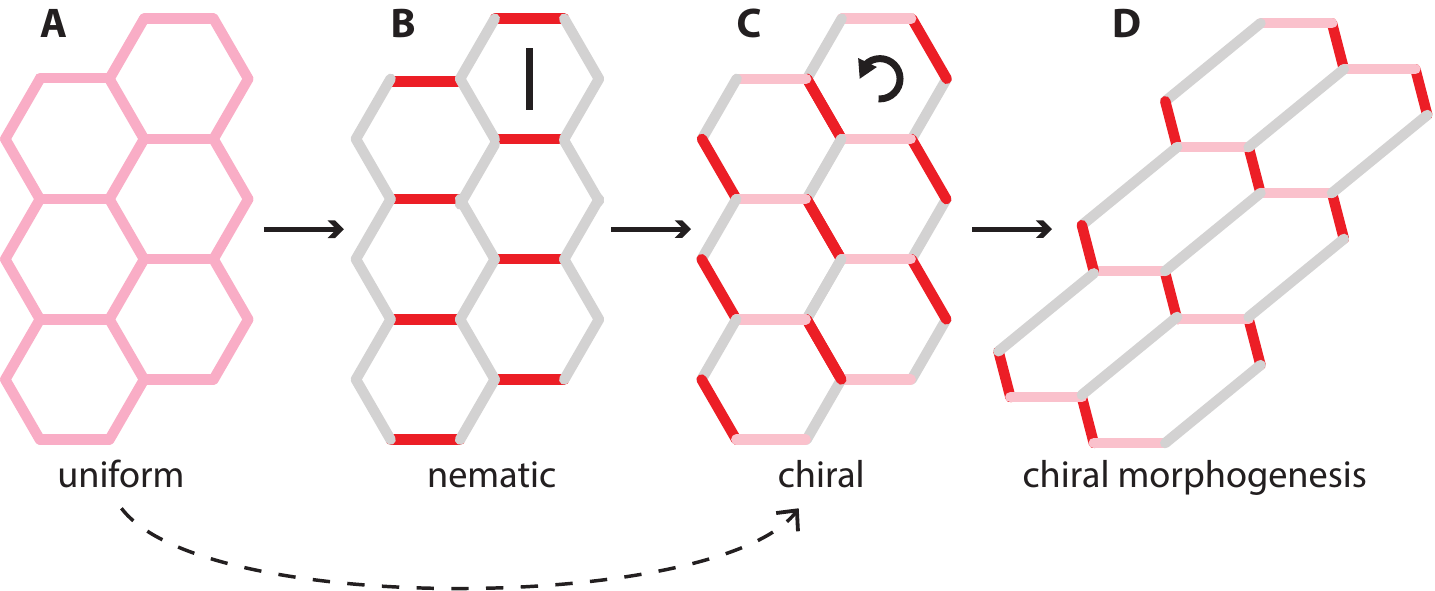}
	\caption{\label{fig1} Chiral morphogenesis driven by spontaneous left-right symmetry breaking. Cells (hexagons) in an epithelial sheet have an initially uniform membrane protein distribution (A, pink) which goes unstable to an intermediate polar (not shown) or nematic (B) state before a secondary instability to a final chiral distribution (C); in certain special cases, a direct transition from isotropic to chiral is possible (dashed line).  Cell and organ shapes then develop a definite handedness in response to chiral protein localization (D).}

\end{figure}

It is only a small exaggeration to view animal development as a progressive breaking of symmetries that sculpts an egg into an elaborate adult form.  One such broken symmetry is the reflection symmetry linking left and right:  Everything from the twist of the \textit{Drosophila} gut to the orientation of the human heart has a definite handedness, but how this handedness is reliably chosen remains incompletely understood.
With few exceptions \cite{BrownWolpert1990, ZhangWolynes2016}, researchers have assumed that chirality at the cellular level and above originates directly from molecular handedness \cite{furthauerGrill2013, naganathanGrill2014, naganathanGrill2016, chen2012, levin2016, tee2015}. In this spirit, considerable effort has gone into exploring mechanisms that might transduce chirality from molecular to cellular scales \cite{speder2007, levin2005, henley2012, inaki2016}. Here, we argue that an alternative scenario, wherein cellular handedness results instead from spontaneous left-right (LR) symmetry breaking, may be equally widespread and is, in particular, a natural explanation for the cell chirality that drives asymmetric morphogenesis in many \textit{Drosophila} organs \cite{inaki2016,taniguchi2011, hatori2014,sato2015NatCom,gonzalez2015}.  In this picture (Fig. \ref{fig1}), the basic mechanism of symmetry breaking does not depend on molecular chirality and does not prefer one handedness over the other. (Molecular handedness can, however, bias this intrinsic tendency towards chirality to ensure that wildtype animals always break symmetry in the same direction.) Specifically, we show that the planar cell polarity (PCP) systems that select a preferred direction in many epithelial tissues \cite{Goodrich2011,Yang2012,Butler2017} are generically also capable of spontaneous LR symmetry breaking.  Thus, establishing cellular chirality may not require any new biological pathways or molecular mechanisms beyond those already characterized for PCP.

Molecular handedness could in principle influence chiral body plans in one of two ways: In the simplest scenario, molecular chirality would propagate directly to larger scales (as it does, for example, in a cholesteric liquid crystal or its active analogs \cite{furthauerGrill2013}); in this case, the magnitude of LR symmetry breaking in the body plan would be proportional to some appropriate measure of molecular handedness, and in particular, the body plan would recover its bilateral symmetry if the chiral molecules driving symmetry breaking were removed.  Alternatively, laterality might be established through the interaction of two distinct systems.  One system would induce an instability in the achiral body plan, leading to spontaneous LR symmetry breaking, but would not encode any preference between left and right.  A second system would then interpret molecular chirality to push the unstable system towards one specific handedness.
Brown and Wolpert \cite{BrownWolpert1990} recognized that the existence of mutations that randomize organ handedness strongly argues for the second scenario; these mutations can be interpreted as knocking out the system that favors one specific chirality while leaving intact the system that drives spontaneous symmetry breaking.  Here, we build on this initial observation by showing explicitly how a symmetry-breaking instability can arise at the cellular (rather than organismal) scale.


It has recently been shown that the handedness of various \textit{Drosophila} visceral organs is evident not only in their overall shape and location within the body, but also in the asymmetry of each of their constituent cells \cite{inaki2016,taniguchi2011, hatori2014,sato2015NatCom,gonzalez2015}.  These organs are formed by monolayer epithelia composed of approximately polygonal cells joined into sheets by specialized junctions (Fig. \ref{fig1}A).  The cellular chirality manifests itself in both protein localization (Fig. \ref{fig1}C) and shape changes (Fig. \ref{fig1}D); for example, in developing male genitalia, Myosin II is localized in a chiral fashion, and cells undergo LR asymmetric T1 topological transitions, leading to clockwise tissue rotation \cite{sato2015NatCom}.  Importantly, mutations have been identified which cause these systems to have bimodal distributions of correct and inverted, but not loss of, handedness \cite{sato2015NatCom,okumura2015}, suggesting that their cell chirality is driven by spontaneous symmetry breaking.  Our goal is to demonstrate that PCP can naturally lead to such an instability.

In the remainder of this paper, we first use general symmetry arguments to determine how membrane proteins can develop chiral polar $(P^*)$ or nematic $(N^*)$ distributions.
Although transitions to the $P^*$ phase have been described in tilted liquid crystal films \cite{selinger1988, selinger1989}, spontaneous symmetry breaking leading to an $N^*$  phase in two dimensions has not to our knowledge been studied before.  To understand what interactions can generate chirality, we next turn to a simple mean-field model, originally developed to described non-chiral PCP \cite{salbreux2012, raymond2014}, and determine which parameters must be varied to achieve chiral configurations. 
Finally, we show that, while it is forbidden in regular hexagonal cells, a continuous transition between isotropic ($I$) and $N^*$ phases is possible in stretched hexagonal cells or when two species of PCP proteins have perfectly symmetrical interactions.

\begin{figure}
	\includegraphics[scale=0.6]{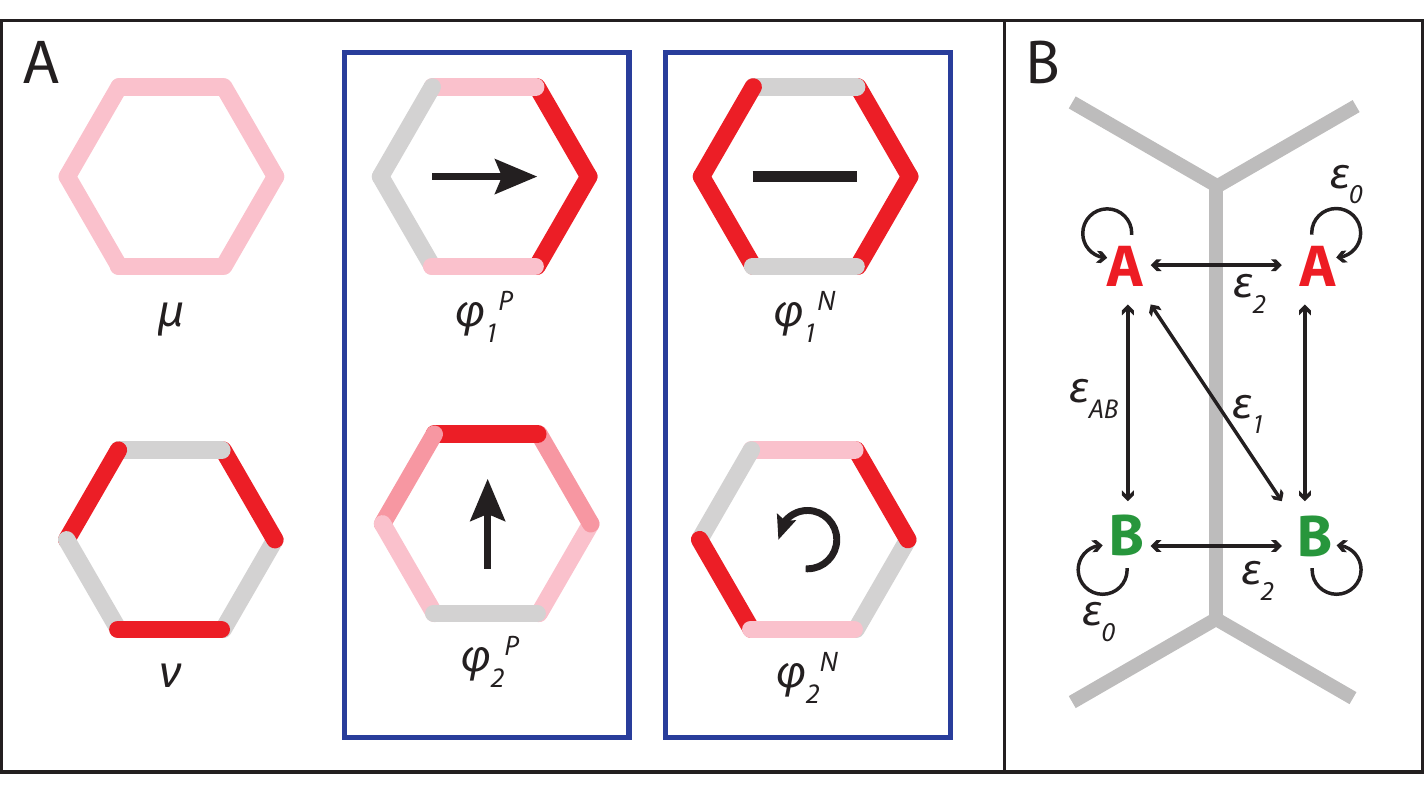}
	\caption{ \label{fig2} (A) Irreducible representations of concentrations on a regular hexagon (point group $D_6$). $\mu$ and $\nu$ are one-dimensional uniform and alternating representations, respectively. The boxed two-dimensional representations transform like polar (left) and nematic (right) order parameters. (B) Mean-field PCP model. $A$ and $B$ are transmembrane proteins residing at cell-cell junctions (grey lines), and the parameters set interaction strengths between: unlike $\left( \varepsilon_{AB} \right)$ and like $\left( \varepsilon_0 \right)$ proteins on the same side of a junction, and unlike $\left( \varepsilon_1 \right)$ and like $\left( \varepsilon_2 \right)$ on opposite sides. Parameters $J_1$, $J_2$, $T$, and $c_{\text{max}}$ are not drawn but are described in the text and in \cite{salbreux2012}.}
\end{figure}

\textbf{General symmetry arguments.}
We begin by working out the generic behavior dictated by symmetry near an instability from an isotropic state \cite{golubitsky1984}. For concreteness, we consider a field of identical hexagonal cells with protein species $A$ and $B$ localized to the cell-cell junctions.  
Because they are based only on symmetry, however, the results of this section apply equally well to systems where cytosolic protein concentrations, cytoskeletal filament orientation, or similar factors play an important role.  We restrict ourselves to the spatially uniform case with an identical protein distribution in every cell, which we assume is completely characterized by the protein concentrations $\mathbf{c} = (c_A^1, c_B^1,\ldots,c_A^6,c_B^6)$ on the 6 hexagon edges; $\mathbf{c}$
 follows the deterministic dynamics $\mathbf{\dot{c}} = \mathbf{f}(\mathbf{c})$.  As usual, we expect that only one mode will initially go unstable, corresponding to some linear combination of protein concentrations (e.g. $A - B$ or $A + B$) whose amplitude varies from edge to edge according to one of the irreducible representations of the hexagon's symmetry group $D_6$ \cite{burak2009} sketched in Fig. \ref{fig2}A.  The first mode, $\mu$, encodes the total protein number and is fixed in this study. The two two-dimensional representations describe polar and nematic order parameters; the choice of basis shown has the advantage that in each case $\varphi_1$ ($\varphi_2$) is even (odd) under reflection about the horizontal axis.  We can thus represent the polar or nematic contribution to the protein concentrations with a complex order parameter $z = x_1 + i x_2$, where $x_1$ ($x_2$) is the coefficent of the corresponding basis vector $\varphi_1$ ($\varphi_2$);
reflection then corresponds to complex conjugation of $z$ and rotation by $\pi/6$ to multiplication by $e^{ i \frac{\pi}{3} }$ ($e^{ i \frac{2\pi}{3} }$) in the polar (nematic) case.

Suppose for the moment that the initial instability is towards nematic order.  Near enough to the instability, other modes can be neglected compared to the slow unstable mode, whose dynamics we can expand in a power series, keeping only terms allowed by symmetry \cite{golubitsky1984}:
	\begin{equation} \label{eqn1}
		\dot{z} = az + 3b\bar{z}^2 + 2c |z|^2 z + \dots.
	\end{equation}
Although the general evolution equation $\mathbf{\dot{c}} = \mathbf{f}(\mathbf{c})$ 
 is not variational, all of the terms in Eq.~\eref{eqn1} can be written as the gradient of an appropriate free energy; at low order, active, far-from-equilibrium systems are thus in this case indistinguishable from equilibrium ones. In fact, it is not until the 4th order in the nematic expansion (7th order in polar) that we get non-variational terms. 

The bifurcation diagram corresponding to Eq.~\eref{eqn1} follows a well-established pattern \cite{golubitsky1984}. Writing $z = r e^{i\theta}$, and keeping higher order terms in $\dot{\theta}$ than in Eq. \ref{eqn1}, we have:
	\begin{align}
		\dot{r} &= ar + 3br^2 \cos(3\theta) +2cr^3 + \dots, \label{eqn2} \\
		\dot{\theta} &= - 3br \sin(3\theta) - dr^3 \sin(3\theta) - 6fr^4 \sin(6\theta) + \dots. \label{eqn3}
	\end{align}
The presence of a quadratic term in Eq. \eref{eqn2} implies that the bifurcation from the isotropic state is subcritical and that the resulting final state is not guaranteed to be perturbatively accessible.  Nonetheless, the generic behavior (which will always hold for sufficiently weakly subcritical bifurcations) can be obtained from the power series expansion.  For small enough $r$, the fixed point always has $\theta = 0$ or $\theta = \frac{\pi}{3}$, depending on the sign of $b$, both of which correspond to the non-chiral configuration $\varphi_1^N$.  The treatment of the polar case proceeds similarly, except that the initial expansion $\dot{z} = a z + 2 c |z|^2 z + \dots$ lacks the quadratic term, implying a supercritical bifurcation; near enough to the bifurcation, the stable states are always purely vertex or edge polarized ($\varphi_1^P$ or $\varphi_2^P$).
Thus, no chiral state is smoothly accessible from isotropic .

Once some order has been established, however, there can be a secondary transition into a chiral state.  As such a transition requires a change in $\theta$, we assume that $r$ at the fixed point is a smooth, single-valued function of $\theta$ and focus only on the latter variable, which, in order to obey the symmetries of Eq. \eref{eqn1}, must satisfy a fixed point equation of the form
	\begin{equation} \label{eqn4}
		\dot{\theta} = 0 = h_3 \sin(3\theta) + h_6 \sin(6\theta)
	\end{equation}
for the nematic case.  Similarly, in the polar case, we have $0 = h_6 \sin(6 \theta) + h_{12} \sin(12 \theta)$ \cite{selinger1988, selinger1989}.  Non-variational terms in the original expansions Eqs. \eref{eqn1}--\eref{eqn3} carry no new $\theta$ dependence, so, here again, equilibrium and far-from-equilibrium systems have the same qualitative behavior.  Eq. \eref{eqn4} can easily be solved graphically.  For $h_6 >0$, there is a discontinuous transition at $h_3 = 0$ between two (non-chiral) nematic states differing by $\pi/3$; for $h_6 < 0$, this transition is replaced by two continuous transitions separated by an intervening chiral phase where $\theta$ varies smoothly between 0 and $\pi/3$ as a function $h_3$.
In exactly the same way, for polar order, $h_{12} > 0$ implies a discontinuous transition between $\varphi_1^P$ (vertex-polarized, $\theta = 0$) and $\varphi_1^P$ (edge-polarized, $\theta = \frac{\pi}{6}$, while for $h_{12} < 0$, the vertex-polarized state undergoes a continuous transition to a chiral state $P^*$ with smoothly-varying $\theta$, which then yields to the edge-polarized state at a second continuous transition as $h_6$ is varied \cite{selinger1988, selinger1989}.





\textbf{Mean-field PCP model.}
The general results just described imply that systems capable of producing polarized protein distributions can always, in principle, also spontaneously break chiral symmetry.  This leaves open the question of whether such symmetry breaking occurs in a biologically realistic parameter regime.  To clarify what interactions must be varied to induce chiral instabilities, we consider a simple mean-field model of PCP \cite{salbreux2012, raymond2014}. This model is governed by a free energy, but, as discussed above, for this particular problem this is sufficient to capture the generic phase behavior even of nonequilibrium systems.  The protein concentrations follow the variational dynamics $\alpha \mathbf{\dot{c}} = -\nabla_{\mathbf{c}} F$, with:
\begin{widetext}
	\begin{equation} \label{eqn5}
	\begin{split}
		F = 
		& \sum_{\substack{\text{cells} \\ \alpha} } \Bigg\{ \sum_{\substack{\text{edges} \\ i\in\alpha} } 
		   \left[ \frac{1}{2}\varepsilon_{AB} l_i c_A^{\alpha,i} c_B^{\alpha,i} - 
		    \frac{\varepsilon_0}{2} l_i  \left( c_A^{\alpha,i} \right)^2 \right] +
		\sum_{\substack{i\in\alpha \\ \beta \text{ shares } \\  i \text{ with } \alpha} }
  		   \left[ - \varepsilon_1 l_i  c_A^{\alpha,i} c_B^{\beta,i}
		    - \varepsilon_2 l_i  c_A^{\alpha,i} c_A^{\beta,i} \right] +
		\sum_{\substack{i,j\in\alpha \\ \langle i,j \rangle} } 
		  \left[ \frac{J_1}{2} \left( c_A^{\alpha,i} - c_A^{\alpha,j} \right)^2 - \right.\\
		   & \left. \frac{J_2}{4} \left( c_A^{\alpha,i} - c_B^{\alpha,j} \right)^2 \right] +
		T \sum_{i\in\alpha} l_i \Big[ c_A^{\alpha,i} \log\left( c_A^{\alpha,i} \right) +
		\left( c_{\text{max}} - c_A^{\alpha,i} \right) \log \left( c_{\text{max}} - c_A^{\alpha,i} \right)  \Big]
		\Bigg\} + 
		\text{(interchange } c_A \leftrightarrow c_B \text{)} .
	\end{split}
	\end{equation}

\end{widetext}
Here, $l_i$ is the length of edge $i$, the $\varepsilon$'s describe interactions between proteins (Fig. \ref{fig2}B), the $J$'s impose a penalty for abrupt concentration jumps, $T$ is a temperature-like parameter, and $c_{\text{max}}$ sets the maximum protein concentration an edge can hold.
The total number of proteins in a cell is fixed and obeys $\sum_{i} l_i c_A^i = 1 - \delta$ and $\sum_{i} l_i c_B^i = 1 + \delta$.
For simplicity, we choose parameters such that the energy (\ref{eqn5}) is symmetric between $A$ and $B$ and adjust the asymmetry between the two species by varying $\delta$; we focus on instability modes that are (approximately) $A-B$ for (non-)zero $\delta$.

%

\begin{figure}
	\includegraphics[width=3.3in]{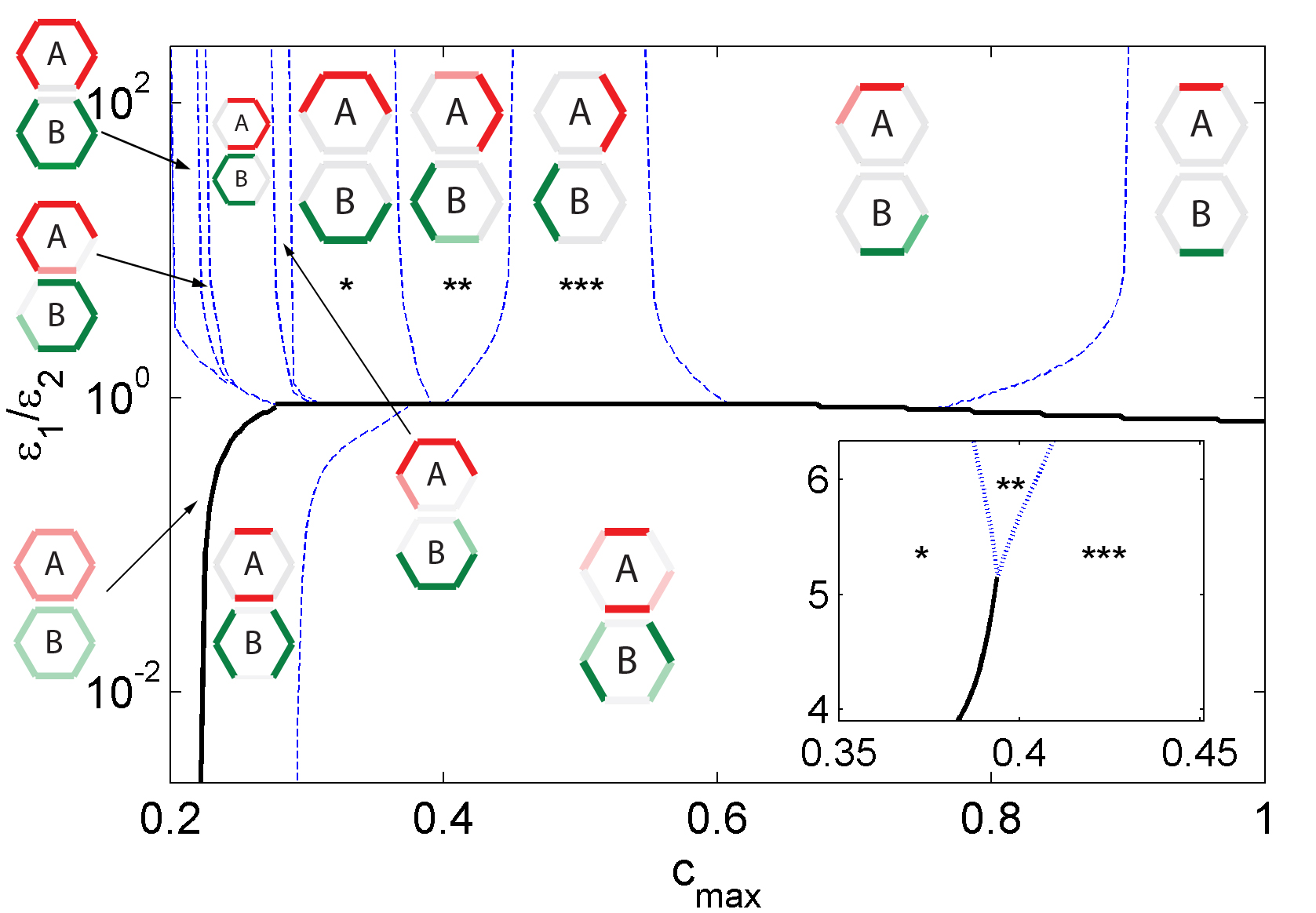}
	\caption{ \label{fig3} Phase diagram of mean-field PCP model showing various polar and nematic chiral phases. Dotted (solid) lines indicate (dis)continuous transitions; schematics show a caricature of the A and B protein distributions in each phase. $\varepsilon_1/\varepsilon_2$ controls the balance between interactions favoring polar and nematic symmetries; $c_{\text{max}}$ sets the maximum number of proteins an edge can hold. Other parameters: $l = \varepsilon_{AB} = 1$, $\varepsilon_0 = 0.02$, $\varepsilon_2 = 0.41$, $J_1 = 0.025$, $J_2 = 0.03$, $T = 0.068$ and $\delta = 0.1$. \textbf{Inset.} Detail at higher $T = 0.3$ showing that a direct, discontinuous transition between vertex and edge-polarized phases is also possible; stars denote corresponding phases in the figure and inset.}
\end{figure}

To explore the behavior of this model, we numerically minimized the free energy \eref{eqn5}.  (The minimization was carried out without imposing any specific symmetry on the solution.)
Fig.~\ref{fig3} shows the resulting phase diagram.  Not suprisingly, the ratio $\varepsilon_1/\varepsilon_2$ (see Fig. \ref{fig2}B and Eq. \eref{eqn5}) governs transitions between phases with predominantly polar or nematic order.  The role of $c_\text{max}$ is more subtle.  Because the quadratic terms in Eq. \eref{eqn5} favor putting as much protein as possible on the same edge, protein will tend to accumulate on as few edges as $c_\text{max}$ allows.  Thus, as $c_\text{max}$ is lowered, relatively high protein concentrations will spread to more edges, potentially creating a chiral state.  For example, if $\varepsilon_1 > \varepsilon_2$, then as $c_\text{max}$ decreases we move from an edge-polarized state where each protein has a high concentration only on one edge, 
through a chiral phase with appreciable but unequal concentrations on two adjacent edges, to a vertex-polarized state with equal concentrations on two edges.  A similar progression from nematic to chiral nematic phases is observed when $\varepsilon_1 < \varepsilon_2$.
The essential physical point is that chiral phases become accessible when the number of edges that the PCP proteins ``want'' to occupy is varied, which in our mean-field model is accomplished by changing $c_\text{max}$.

\textbf{Direct $\mathbf{I \rightarrow N^*}$ transitions.} Although direct, continuous transitions from isotropic to chiral are normally prohibited by the hexagon's $D_6$ symmetry, such transitions are possible when: (1) the cells are stretched so that the symmetry is lowered to $D_2$, or (2) protein species $A$ and $B$ have perfectly symmetrical interactions.

\begin{figure}
	\includegraphics[width=3in]{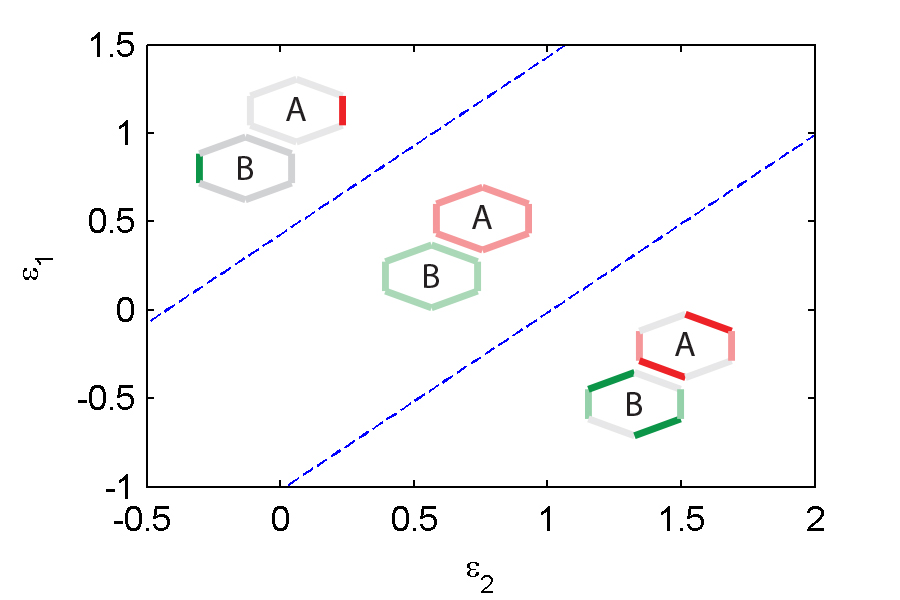}
	\caption{ \label{fig4} Phase diagram of mean-field PCP model on a stretched hexagon, showing a continuous $I$ to $N^*$ transition that does not exist for unstretched hexagons.  Long and short hexagon edges have lengths 1.3 and 0.7; $J_2 = -0.8$; $c_\text{max} = 0.7$; other parameters as in Fig. \ref{fig3}. }
\end{figure}

In the first case, stretching regular hexagons splits the polar and nematic representations into four different one-dimensional representations corresponding to the basis vectors shown in Fig. \ref{fig2}A. Of these, $\varphi_2^N$ is evidently chiral, and expanding its dynamics in analogy to Eq. \eref{eqn1}, one quickly sees that reflection symmetry prohibits a quadratic term, implying a possible supercritical bifurcation.  Fig. \ref{fig4} shows that the mean-field model on stretched hexagons indeed exhibits a continuous $I$ to $N^*$ transition.  Epithelial tissues are frequently under tension along some body axis, allowing them break LR symmetry without passing through any intermediate states.

In the second case, imposing symmetry under exchange of $A$ and $B$ forces the dynamics of the nematic $A-B$ mode, Eq. \eref{eqn1}, to be invariant under $z \mapsto -z$, excluding all even order terms on the righthand side.  In the absence of the quadratic term $3 b \bar{z}^2$, the system is generically capable of a supercritical bifurcation.  Interestingly, the dynamics of the nematic mode with exchange symmetry and of the polar mode 
are formally identical, but the physical intepretation is very different.  Depending on the sign of the coefficient of $\bar{z}^5$, the state immediately above the bifurcation will have either $\theta = n \pi/3$ or $\theta = \pi/6 + n \pi/3$ ($n \in \mathbb{Z}$).  In the polar case, these correspond to vertex and edge polarization, but when $z$ instead describes a nematic, $\theta = \pi/6 + n \pi/3$ is a chiral state.  Although it seems unlikely that a living system would ever have perfect $A \leftrightarrow B$ exchange symmetry, such a situation might be realized in an \textit{in vitro}, biomimetic system, where $A$ and $B$ could correspond, for example, to opposite enantiomers of otherwise identical molecules.
Weakly breaking $A \leftrightarrow B$ symmetry preserves the chiral phase, but introduces a narrow intervening non-chiral region between $I$ and $N^*$.

\textbf{Discussion.}  We have shown that PCP represents a general route to spontaneous LR symmetry breaking at the cellular level.  Chiral symmetry is generically broken via a secondary, continuous transition from a state that already exhibits polar or nematic order, but direct transitions from an isotropic to a handed protein distribution are possible when the bifurcation to the nematic state is sufficiently strongly subcritical, when cells are subject to uniaxial stretching, and when PCP protein interactions are invariant under exchange of the protein species $A$ and $B$.  This last case represents a new class of transition, distinct from those previously described in liquid crystals and other soft matter systems, that might be experimentally accessible \textit{in vitro}.  Indeed, although we have focused primarily on living epithelia, our models apply equally well to biomimetic systems composed of arrays of vesicles or liquid drops functionalized with appropriate adhesion molecules \cite{Pontani2016}.  In either case, a key quantity controlling the transition from achiral to chiral is the number of polygon edges with high concentrations of a given protein species.  In our mean-field model, this variable
is governed by the interplay between the total protein number and the maximum concentration $c_\text{max}$ on a given edge, 
but more generally it could also depend, for example, on an appropriate correlation length for protein interactions along the cell perimeter \cite{burak2009}.

An important implication of our results is that spontaneous LR symmetry breaking can be explained without invoking any new, uncharacterized proteins or interactions:  The PCP pathways commonly found in epithelial tissues are sufficient to produce chiral protein localization by appropriate parameter modulation, 
which could be accomplished by up or down-regulation of the expression of various PCP components, by changes in protein activity or affinity induced by covalent modification or small molecule binding, or even by mechanical changes to cell shape.  
Once a chiral protein distribution has been established, it can preferentially shrink and remodel junctions in an LR asymmetric manner and thus drive chiral tissue and organ morphogenesis.

Handedness appears to arise in different organisms by a variety of distinct mechanisms \cite{Vandenberg2013,McDowell2016}.  In some contexts, PCP contributes to chiral symmetry breaking not directly, through chiral protein localization, but indirectly, by establishing an axis relative to which other chiral activities can then be expressed \cite{Zhang2009,Antic2010,Song2010}.  More relevant to our findings are several \textit{Drosophila} systems where asymmetric organ morphogenesis is believed to be driven by cell chirality:  In the testis and both the embryonic and adult gut, looping of tubular organs is preceded by the appearance of LR asymmetry in protein localization and in cell shape and movement \cite{taniguchi2011, hatori2014,sato2015NatCom,gonzalez2015}.  Importantly, several mutants in these systems give rise to bimodal distributions of wildtype and reversed handedness, as expected for spontaneous symmetry breaking.  These include knockouts of the unconventional \textit{myosin ID} \cite{Hozumi2006,geminard2014,hegan2015}, which leads to partially penetrant inversion of the embryonic hindgut \cite{okumura2015}, and of the myosin regulatory light chain \textit{spaghetti squash}, which
causes a fraction of cells in male genitalia to reverse their orientation \cite{sato2015NatCom}.  We predict that partially knocking down these factors would induce a change in the ratio of normal and inverted phenotypes but no loss of chirality.  In contrast, if PCP is responsible for the underlying symmetry breaking, then impairing it should cause a completely achiral phenotype.  This is indeed what is seen in the adult hindgut, where PCP appears to ``memorize'' a handedness initially imposed by a small organizer region that subsequently detaches from the gut \cite{gonzalez2015}.  Our results provide a straightforward mechanism for this phenomenon.  Interestingly, loss of the organizer also destroys chirality; we expect in this case that, even if global organ looping is perturbed, individual cells or groups of cells will still show chiral protein localization.  In fact, an obvious prediction of our model is that PCP protein concentrations or activities should adopt a handed pattern in epithelia with cell chirality.  Although ideally this would be verified by direct observation, such asymmetries are not always easy to detect \cite{gonzalez2015}; alternative tests could involve knocking down PCP proteins in clones and observing the shape and orientation of nearby cells.

We have limited ourselves to deterministic, mean-field dynamics in spatially uniform systems, where we find the same qualitative behavior near to or far from equilibrium.  It is well-known that fluctuations in two dimensions can lead to significant deviations from mean-field behavior (even when the broken symmetry is not continuous \cite{wu1982}).  We thus expect that some of our results will break down in the presence of enough noise and on long enough length scales, even as there is a good chance that they remain valid for biological tissues with only hundreds or thousands of cells.  Even at the mean-field level, spatially-modulated phases could also intrude on the phase diagram \cite{selinger1993}; moreover, once gradient terms are added to the dynamics, the behaviors of active and passive systems are expected to diverge \cite{marchetti2013}.  Finally, most real cell packings deviate from the perfect hexagons we have assumed, and strong enough disorder is likely to modify the bifurcations we describe.  At the same time, equilibrium phase transitions often survive weak disorder \cite{Lubensky1979}, and a similar robustness has recently been observed in models of PCP \cite{Yamashita2016}, suggesting that our conclusions will remain at least qualitatively correct for many epithelia.

This work was funded by NSF grants DMR-1056456 and IOS-1353914.  We are grateful for helpful conversations with Guillaume Salbreux.

\end{document}